\documentclass[12pt,a4paper]{article}
\usepackage{a4wide}
\usepackage{graphicx}
\usepackage{fullpage}
\begin{document}
\title{Production of QED pairs at small impact parameter
in relativistic heavy ion collisions}
\author{Kai Hencken, Gerhard Baur, Dirk Trautmann}

\maketitle

\begin{abstract}
The STAR collaboration at RHIC is measuring the production of electron-positron
pairs at small impact parameters, 
larger than but already close to the range, where the ions interact strongly
with each other.
We calculate the total cross section, as well as, differential distributions
of the pair production process with the electromagnetic 
excitation of both ions in a semiclassical approach and within a lowest
order QED calculation. We compare the distribution of electron and positron
with the one coming from the cross section calculation without restriction 
on impact parameter. Finally we give an outlook of possible results at the LHC.
\end{abstract}

\section {Introduction}
Pair production in relativistic heavy ion collisions has attracted 
interest in the past mainly due to the fact that the strong fields allow for
multiple pairs to be produced. At impact parameter of the order of twice
the nuclear radius, but still larger than this, so that the two ions do not 
interact hadronically with each other (that is the regime of the so-called 
``ultraperipheral collisions'' UPC),
the total pair production multiplicity is found to be
about $1.5$ for AuAu collisions at RHIC and about $3.9$ for PbPb collisions
at the LHC. These results are based on a lowest order QED calculation
\cite{HenckenTB95b} and one might suspect that at these small impact parameters
the strong fields of the two ions do lead to higher order corrections.

\begin{figure}[tb]
     \centering
     \includegraphics[width=6cm]{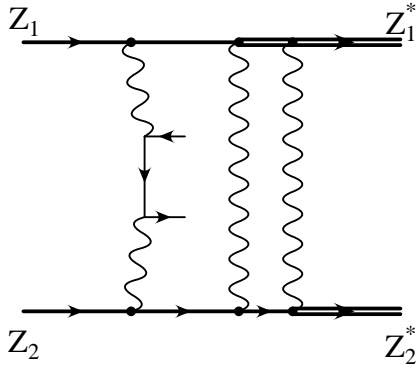}
\caption{The pair production together with the electromagnetic excitation
of both ions, predominantly to the giant dipole resonance (GDR) is shown
as one typical Feynman diagram.
The process in lowest order involves at least the exchange of four photons
(many more ``soft Coulomb photons'' are exchanged as well). Due to this
the process predominantly occurs at small impact parameter, where the 
electromagnetic fields are strong, in contrast to the unrestricted cross section
(without electromagnetic excitation of the ions), which has contributions 
coming also from large impact parameters.
}
\label{fig:epemGDR}
\end{figure}

Therefore it is of interest to measure electron-positron pairs
produced in such collisions and compare their distribution with 
theoretical predictions, e.g., in lowest order QED. The STAR collaboration
has recently measured $e^+e^-$ pairs in 
collisions, which were selected by a trigger, looking for the simultaneous 
excitation of the two ions (mainly to the GDR) in addition to the pair 
production
process \cite{Meissner:2003me,Morozov:2003wk,Klein:2003rx}, 
see Fig.~\ref{fig:epemGDR}. 
Such an event is characterized 
by the subsequent emission of one or only a few neutrons, which are then
detected in the forward ZDC (``Zero Degree
Calorimeter''). 
This tagging was first proposed in order to study vector meson production
\cite{Baltz:2002pp,KleinN00,KleinN99}. Mutual excitation of the two ions
is also used for the luminosity measurement at RHIC \cite{Baltz98,Chiu:2001ij}.

As the average impact parameter in such 
collisions is only about \cite{Baur:2003ar}
\begin{equation}
\overline b = {\int d^2b\  b P(b) \over 
\int d^2b P(b)}
\approx
\frac{8R_a}{3} \approx 19\mbox{~fm},
\end{equation}
one may expect strong field effects to be present for the pairs.

Due to the design of the STAR detector
only electrons and positrons having a transverse momentum
$p_t>65$~MeV/$c$ and being emitted with a rapidity $|y|<1.15$ can
be detected. As most of the pairs produced in ultraperipheral
collisions (UPCs) are emitted with energies of the order of a few $m_e
c^2$ and at small angles, such a measurement can only look at the tails
of the distribution of the pairs.
Even this momentum range was only possible due to a lowered magnetic field
in the STAR detector.

One approach to the calculation of this process is the use of the impact
parameter dependent equivalent photon approximation 
\cite{Baur90d,BaurF90,GreinerVRS91,CahnJ90}. The additional electromagnetic
processes are easily incorporated in this semiclassical approach and the cross 
section can be expressed as
\begin{equation}
\frac{d^6\sigma_{e^+e^-,2GDR}}{d^3p_+d^3p_-} = \int d^2b P^2_{GDR}(b) 
\frac{d^4L}{d^2b d\omega_1 d\omega_2} 
\frac{d^6\sigma_{\gamma\gamma\rightarrow e^+,e^-}(\omega_1,\omega_2)}
{d^3p_+ d^3p_-}.
\end{equation}
where $\sigma_{\gamma\gamma\rightarrow e^+,e^-}$ denotes the cross section
for real photons and $d^4L/d^2b d\omega_1 d\omega_2$ the impact parameter 
dependent photon-photon luminosity
(for the details of the photon-photon luminosity, see Sec.~2.7 of 
\cite{BaurHT98} and references therein).
One difficulty in this approach is the correct choice of the cutoff parameter 
present in the expression for the photon spectrum. This is especially difficult
due to the smallness of $m_e$, which is 
much smaller than the ``usual cutoff'' imposed
on the maximal transverse momentum of the photon from the elastic form factor
of the ion, which is given by $1/R_A\approx 80$~MeV. It was found that the 
total cross section for electron-positron pair production is only reproduced
with a cutoff chosen around $m_e$. On the other hand it was also found that 
neither a
choice of $m_e$ nor of $R_A$ is able to predict the total probability at
impact parameter smaller than the Compton wave length $\lambda_c=386$~fm 
\cite{HenckenTB95b}. For a discussion about the choice of the impact parameter
see, e.g., \cite{BudnevGM75}. In addition in the usual semiclassical approach 
the transverse momentum distribution of the photons is integrated over. In 
order to get, for example, the transverse momentum distribution of the pair, 
one needs to take this momentum distribution into account to get the correct
final result. For a possible approach taking this into account from first 
principles see \cite{VidovicGB93,BaurB93}.

In Section~\ref{sec:theory} we show how our calculation is done
in lowest order QED and in the semiclassical approximation. This is then
used in Sec.~\ref{sec:rhic} to calculate total cross sections,
differential distributions for RHIC and also for possible LHC conditions.
The comparison with the experimental results has been done in the
meantime and will be presented by the STAR collaboration in
another publication \cite{Klein04}.

\section{Calculation of pair production and nuclear excitation in lowest
order QED}
\label{sec:theory}

The STAR experiment at RHIC measures the pair production cross
section together with the double 
electromagnetic excitation in both ions, see 
Fig.~\ref{fig:epemGDR}. 
In order to incorporate the experimental conditions in the theoretical 
calculation, it is most appropriate to work in the semiclassical approach.
Using the fact, that in this approach the 
probabilities of the individual processes factorize and are given by
the product of the individual probabilities (for a theoretical description
of this approach, see \cite{Baur:2003ar}) we can write the cross 
section for this process as
\begin{equation}
\frac{d^6\sigma_{e^+e^-,2GDR}}{d^3p_+d^3p_-} =
2 \pi \int_{b_{min}}^{\infty} b db P_{GDR}^2(b) \frac{d^6P(b)}{d^3p_+ d^3p_-}.
\label{eq:sigmasemicl}
\end{equation}
The minimum impact parameter was chosen to be $b_{min}=2R_a\approx 14$fm,
where we assume the nuclei to touch, that is, interact hadronically with each 
other.
The use of the semiclassical description in this case is not only justified
due to the strong Coulomb interaction between the two ions (leading to a large
number of ``soft'' Coulomb photons exchanged between them), but also simplifies
the calculation of this higher order process (pair production plus two GDR
excitation processes) considerably.

Following \cite{BertulaniB88} the probability for GDR excitation in one ion
is to a good approximation given as
\begin{equation}
P_{GDR}(b) = S/b^2
\label{eq:pgdrb2}
\end{equation}
with 
\begin{equation}
S = \frac{2\alpha^2 Z^3 N}{A m_N\omega} \approx
5.45 \times 10^{-5} Z^3 N A^{-2/3} \mbox{fm}^2.
\label{eq:Sgdr}
\end{equation}
where $m_N$ denotes the nucleon mass, and the neutron-,
proton-, and mass-number of the ions are 
$N,Z,$ and $A$ respectively (we consider only symmetric collisions here,
the calculation can trivially be extended to incorporate also asymmetric 
systems). The excitation probability is inversely proportional 
to the energy $\omega$ $(\approx 80\mbox{MeV}A^{-1/3})$ of the GDR state. 
Neutrons are not only emitted from the GDR excitation but are also coming
from higher excited states \cite{Baltz98,BaltzRW96,BaltzS98}. 
These can be taken into account approximately by increasing $S$ 
accordingly. As this does not change the $1/b^2$ behavior for the small
impact parameter, we are interested in, and only leads to a rescaling of the
total cross section, not the form of the differential distributions, 
we have used in our calculation
the more simple value of $S$ in Eq.~(\ref{eq:Sgdr}).
Of course one needs then to include
in addition the different decay channels into one, two, etc. 
neutrons \cite{Pshenichnov:2001qd,Pshenichnov:1999hw,Pshenichnov98},
as well as, the fact that the probability for GDR excitation is 
already quite large (about 0.5 for $b_{min}$) so that multiphoton excitation
mechanisms need to be included as well. Assuming a Poisson distribution
for the different (independent) excitation processes, one would need to
replace $P_{GDR}(b)$ then with \cite{BaltzRW96}
\begin{equation}
P(b) = 1 -\exp(P_{GDR}(b)).
\end{equation}
One sees that the multiphoton excitation tends to reduce again the 
probability (it has to stay below the unitarity limit of 1). 
In the appendix we show how a more complex $b$ dependence than
a simple $1/b^2$ for $P_{GDR}(b)$ can be calculated within the same approach.
Therefore an improved calculation can be done in this way.
On the other hand for the current experimental accuracy the simplified 
approach seems to be appropriate.

With this the cross section for pair production together with the GDR 
excitation of both ions is given by
\begin{eqnarray}
\frac{d^6\sigma_{e^+e^-,2GDR}}{d^3p_+d^3p_-} &=& 
2\pi \int_{b_{min}}^\infty b db P_{GDR}^2(b) 
2\pi \int q dq \frac{d^6\hat P(q)}{d^3p_+d^3p_-} J_0(q b)\\
&=& (2\pi)^2 S^2 \int q dq \frac{d^6\hat P(q)}{d^3p_+d^3p_-}
\int_{b_{min}}^\infty b db \frac{J_0(q b)}{b^4}\\
&=& (2\pi)^2 S^2 \int q dq \frac{d^6\hat P(q)}{d^3p_+d^3p_-} 
\int_{b_{min}}^\infty \frac{db}{b^3} J_0(q b)
\end{eqnarray}
where we have introduced the two-dimensional Fourier transform of the
impact parameter dependent probability for pair production 
$d^6P(b)/d^3p_+d^3p_-$ as
\begin{equation}
\frac{d^6P(b)}{d^3p_+ d^3p_-} = \int d^2 q \exp(i \vec q \vec b) 
\frac{d^6\hat P(\vec q)}{d^3p_+ d^3p-}
= 2\pi \int q dq \frac{d^6\hat P(q)}{d^3p_+d^3p_-} J_0(q b)
\end{equation}

We rewrite the integral over $b$ in dimensionless units as
\begin{equation}
\int_{b_{min}}^\infty \frac{db}{b^3} J_0(q b) =
q^2 \int_{q b_{min}}^\infty \frac{dx}{x^3} J_0(x) =: q^2 I_3(q b_{min}).
\label{eq:iqb}
\end{equation}

\begin{figure}[tb]
     \centering
     \includegraphics[width=3cm]{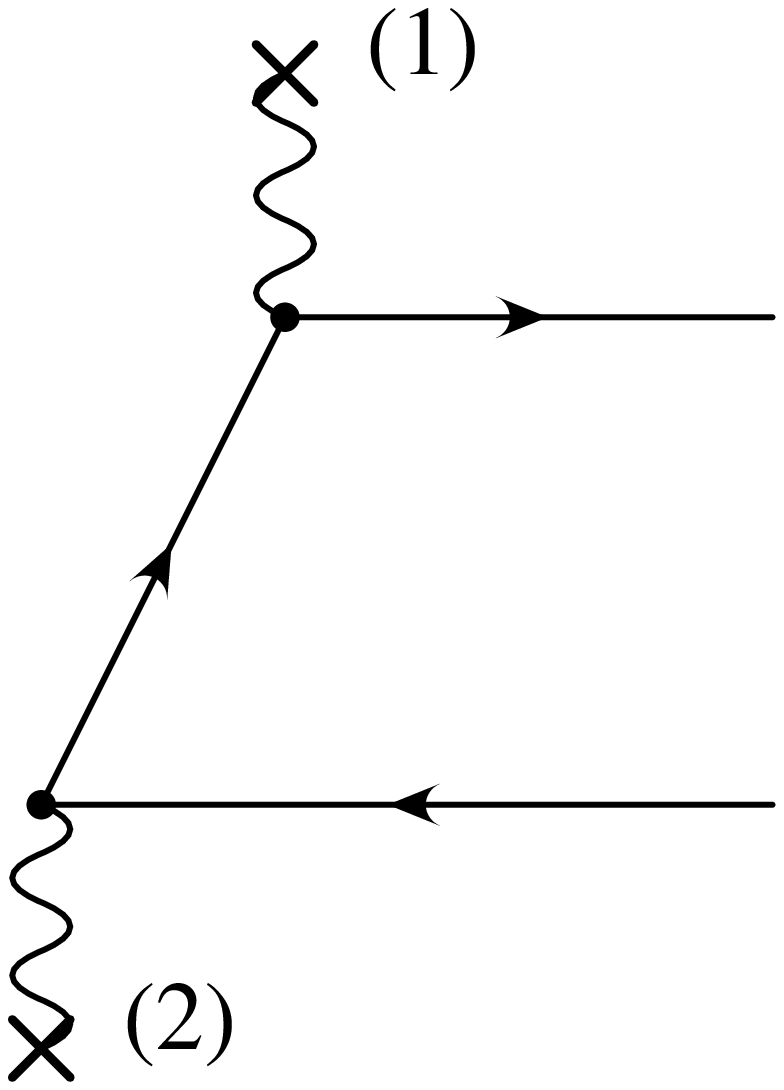}
     \includegraphics[width=3cm]{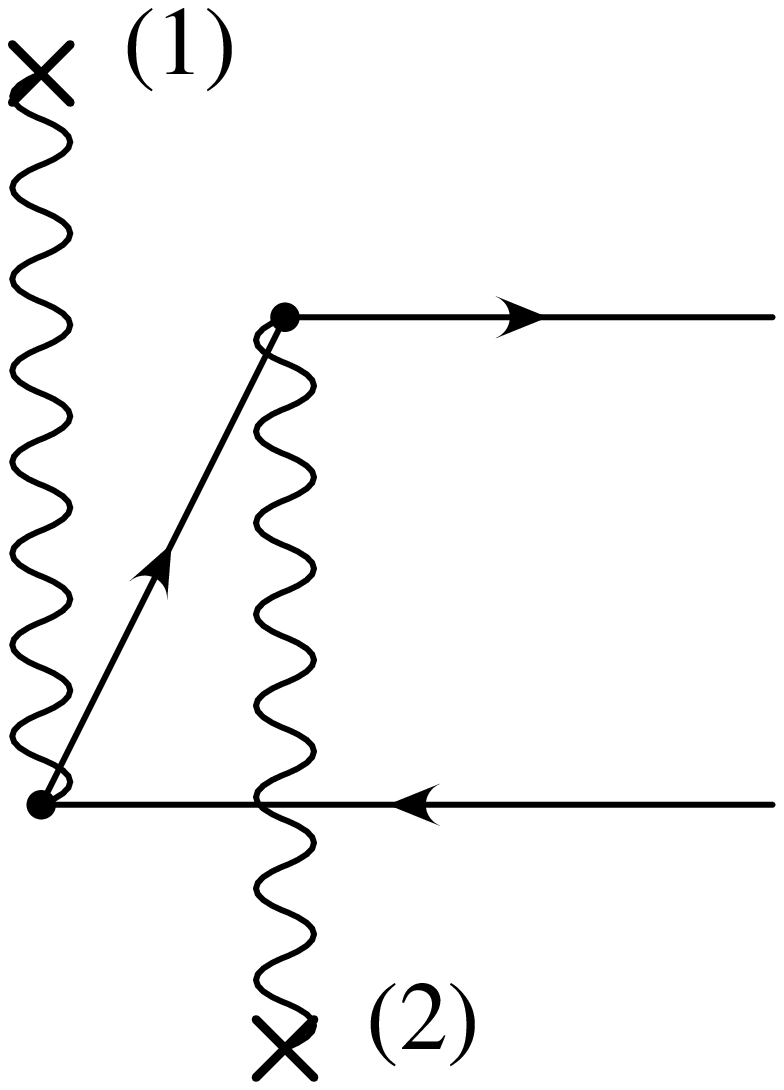}
\caption{The two Feynman diagrams contributing to pair production in lowest 
order QED are shown. The crossed denote the coupling to the (external) Coulomb
field of one of the ions.}
\label{fig:fd}
\end{figure}

Following the derivation of \cite{HenckenTB95b,AlscherHT97}, one can calculate 
the two-dimensional Fourier transform $\frac{d^6P(\vec b)}{d^3p_+ d^3p_-}$
in lowest order QED. 
The two Feynman diagrams for this process in the semiclassical approximation 
are shown in Fig.~\ref{fig:fd}. One gets the differential probability as
\begin{eqnarray}
\lefteqn{\frac{d^6\hat P(q)}{d^3p_+d^3p_-} = (Z\alpha)^4 
\frac{4}{\beta^2\gamma^4} 
\frac{1}{(2\pi)^6 2 \epsilon_+ 2 \epsilon_-}
 \int d^2q_1 [N_0 N_1 N_3 N_4]^{-1} }\nonumber\\
&&\times \mbox{Tr} \biggl\{(\not\!p_- + m) \left[ 
  N_{2D}^{-1} \not\!\!u^{(1)}(\not\!p_- - \not\!q_1 + m)\not\!\!u^{(2)}
  + N_{2X}^{-1} \not\!\!u^{(2)} (\not\!q_1 - \not\!p_+ + m) \not\!\!u^{(1)}
  \right] \nonumber\\
&&\times (\not\!p_+ - m) \left[
  N_{5D}^{-1} \not\!\!u^{(2)}(\not\!p_- - \not\!q_1' + m)\not\!\!u^{(1)}
  + N_{5X}^{-1} \not\!\!u^{(1)} (\not\!q_1' - \not\!p_+ + m) \not\!\!u^{(2)}
  \right] \biggr\} 
.
\end{eqnarray}
with
$$
N_0 = - q_1^2,\quad N_1 = - (q_1^2 - (p_+ + p_-))^2,
$$
$$ N_3 = - (q_1 + q)^2, \quad N_4 = - [q_1 + (q - p_+ - p_-)]^2,$$
$$N_{2D} = -(q_1 - p_-)^2 + m^2, N_{2X} = - (q_1 - p_+)^2 + m^2,$$
$$N_{5D} = -[q_1 + (q - p_-)]^2 + m^2, N_{5X} = -[q_1 + (q - p_+)]^2 + m^2,$$
with the longitudinal component of $q_1$ given by
$q_{1 0} = \frac{1}{2} \left[ (\epsilon_+ + \epsilon_-)
  + \beta (p_{+ z} + p_{-z})\right]$,
$q_{1z} = \frac{1}{2 \beta} \left[ (\epsilon_+ + \epsilon_-)
  + \beta (p_{+z} + p_{-z})\right] = \frac{1}{\beta} q_{10}$
and $u^{(1,2)}=\gamma (1,0,0,\pm \beta)$ are the four velocity of the two ions,
$\gamma$ and $\beta$ the Lorentzfactor and velocity of each ion in the center of mass frame. We have included in addition a nuclear form factor $F(q)$. We 
choose for ease of computation in our case a monopole form factor 
of the form
\begin{equation}
  F(q)= \frac{\Lambda^2}{\Lambda^2 - q^2}
 = \frac{\Lambda^2}{\Lambda^2+Q^2}
\end{equation}
where $\Lambda^2=\frac{6}{<R^2>}$ is set to about 80~MeV in order to reproduce 
the rms radius of the ion. This leads in the terms $N_0^{-1}$, $N_1^{-1}$, 
$N_3^{-1}$ and $N_4^{-1}$ to a replacement of the term $1/q^2$ by $F(q)/q^2$. 
The integration over $d^2q_1$ can be done analytically, using the
usual tricks for Feynman integrations in two dimensions. For details
of this, we refer the reader to \cite{HenckenTB95b}. 

The integral $I_3(z=q b_{min})$ in Eq.~(\ref{eq:iqb}) can be solved 
analytically and calculated easily, as is shown in the appendix.

Finally we make the integral over $dq$ dimensionless to get
\begin{equation}
\frac{d^6\sigma_{e^+e^-,2GDR}}{d^3p_+d^3p_-} = (2 \pi)^2 \frac{S^2}{b_{min}^4}
\int z^3 dz \frac{d^6\hat P\left(\frac{z}{b_{min}}\right)}{d^3p_+d^3p_-} I(z).
\label{eq:sigsb}
\end{equation}

For the unrestricted differential cross section (that is without triggering on
the additional electromagnetic excitations of the ions), we integrate
over $\vec b$ without the factor $P^2_{GDR}(b)$ in Eq.~(\ref{eq:sigmasemicl}).
As the contribution coming from 
$b<b_{min}$ is small, we have extended the integration over all $b$.
\begin{eqnarray}
\frac{d^6\sigma_{e^+e^-}}{d^3p_+d^3p_-} &=& \int d^2b\  
\frac{d^6P(b)}{d^3p_+d^3p_-}
\nonumber\\
&=& \int d^2b\  d^2 q\  
  \tilde \frac{d^6\hat P(q)}{d^3p_+d^3p_-} \exp(i \vec q \vec b) 
  \nonumber\\
&=& (2 \pi)^2 \int d^2q\  \delta(\vec q)
\frac{d^6\hat P(q)}{d^3p_+d^3p_-} \nonumber\\
&=& (2 \pi)^2 \frac{d^6\hat P(0)}{d^3p_+d^3p_-}.
\label{eq:sig}
\end{eqnarray}
This approach was pursued in \cite{AlscherHT97} and total and differential
cross sections were calculated.
For the unrestricted cross sections we do not take a nuclear form
factor of the two ions into account, as pair productions occurs predominantly 
at large impact parameter and for small $q^2$ of the two photons. 

Whereas for the unrestricted
cross section only the value of $\hat P(q)$ for $q=0$ is 
needed, here our expression is a folding over a
range of $q$ given in terms of $1/b_{min}$. In order to compare the 
differential distributions in both the restricted and in the unrestricted case
we have also made calculations of the unrestricted cross section with the
same kinematical restrictions as in the case with GDR excitations.

The expression of Eq.~(\ref{eq:sigsb})
is in a form, which can be evaluated using a Monte Carlo 
integration for both the integration over $z$, as well as, the
six-dimensional integration over $p_+$ and $p_-$ at the same time.
For this we have used VEGAS \cite{Lepage78}.
Both $P(q)$ and $J_0^{int}$
are oscillatory functions, having both positive and negative 
values, which could lead to cancellations. Looking at $z^3 I(z)$ together 
with the result of $P(z/b_{min})$ one sees that the
integrand falls off for large $z$, that is for large $q$.
It is found that the main contribution comes from the
region around $z=2$ and that the contribution from the negative part at larger
$z$ are suppressed. The integration will have positive
and negative contributions but the cancellations between them are not severe.
With the help of VEGAS we can get the total cross section and also
differential cross sections by binning the differential results.

\section{Results}
\label{sec:rhic}

We have made calculations of the total cross section and 
differential distributions of the electron, the positron and the pair 
including the experimental restrictions at STAR.
The integration over $b$ (or equivalently $q$)
is incorporated into the Monte Carlo integration. Another strategy would be to calculate
$d^6\hat P(q)/d^3p_+d^3p_-$ for different values of $q$ and fixed values of 
$p_+,p_-$ and do a Bessel transform in each case. For the total cross section,
that is $\hat P(q)$, this can be done and was done as an 
independent check of our approach. To obtain differential cross sections, this 
approach is rather cumbersome.

\begin{figure}[tb]
     \centering
     \includegraphics[width=7cm]{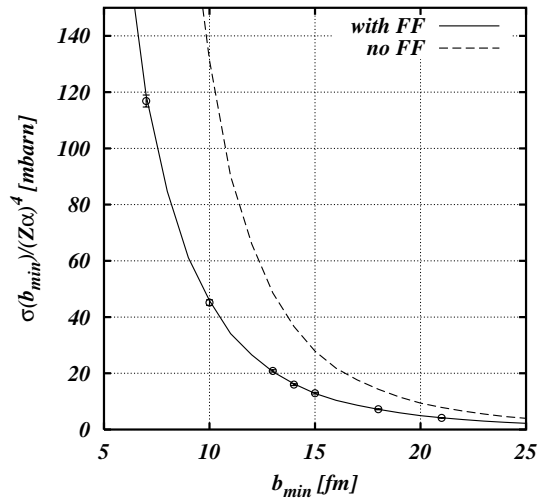}
\caption{The total cross section including kinematical restrictions
of the STAR experiment are shown as a function of the minimal allowed
impact parameter $b_{min}$. Calculations with and without a form factor 
for the nucleus are shown. The lines were calculated by calculating $P(b)$
first and integrating then over $b$. These lines are in perfect agreement
with a similar approach, where $P(q)$ is calculated first and then integrated
over $q$. The circles correspond to the results of our Monte Carlo approach, 
where, as explained in the text, both the integration over $q$ and over $p_+$
and $p_-$ are done within the Monte Carlo integration routine.}
\label{fig:total}
\end{figure}

In a first step, we have calculated the total cross section as a function of 
$b_{min}$, the minimum impact parameter, by using three different approaches: 
We can calculate $P(b)$ directly for the pair production process and 
integrate numerically over $b$. Second we can start from $\hat P(q)$ 
directly and do the integration over $q$ numerically via the Fourier 
transformed of $P_{GDR}(b)$, that is, using $I(z)$.
Finally we have done the calculation with the integration over $q$, that is 
$z$, directly with the Monte Carlo integration. In all three cases we have 
restricted the phase-space integration over the momenta of electron and 
positron according to the experimental conditions of STAR: $p_t>60$~MeV$/c$
and $|y|<1.15$ for each lepton. The results are shown in 
Fig~\ref{fig:total}. The lines correspond to calculations with and without a 
monopole form factor for the nucleus, showing that the incorporation of a form
factor is important. 
All three results agree quite well with each other, showing that our 
approach is working well.

\begin{figure}[tb]
     \centering
     \includegraphics[width=7cm]{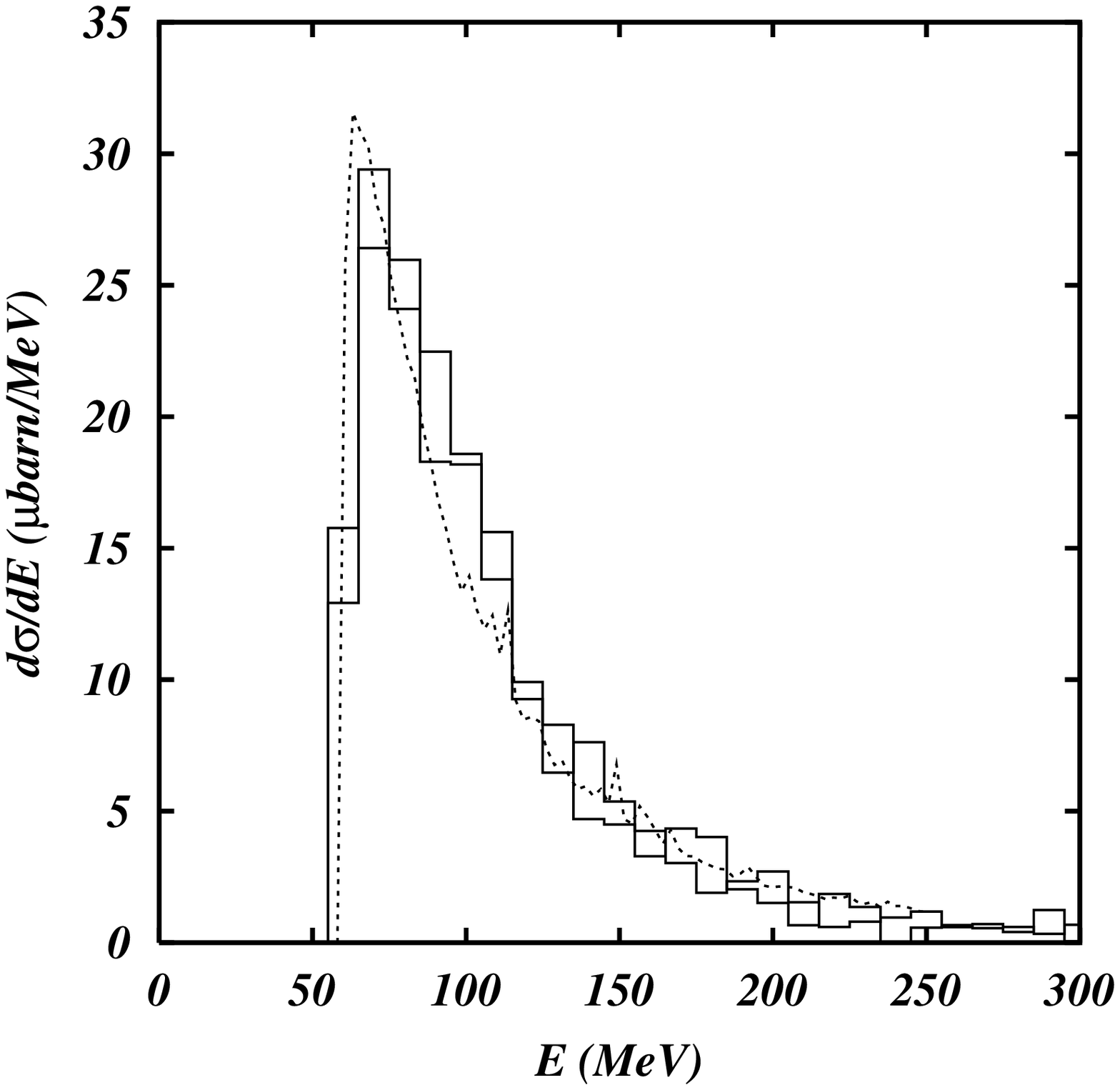}
     \includegraphics[width=7cm]{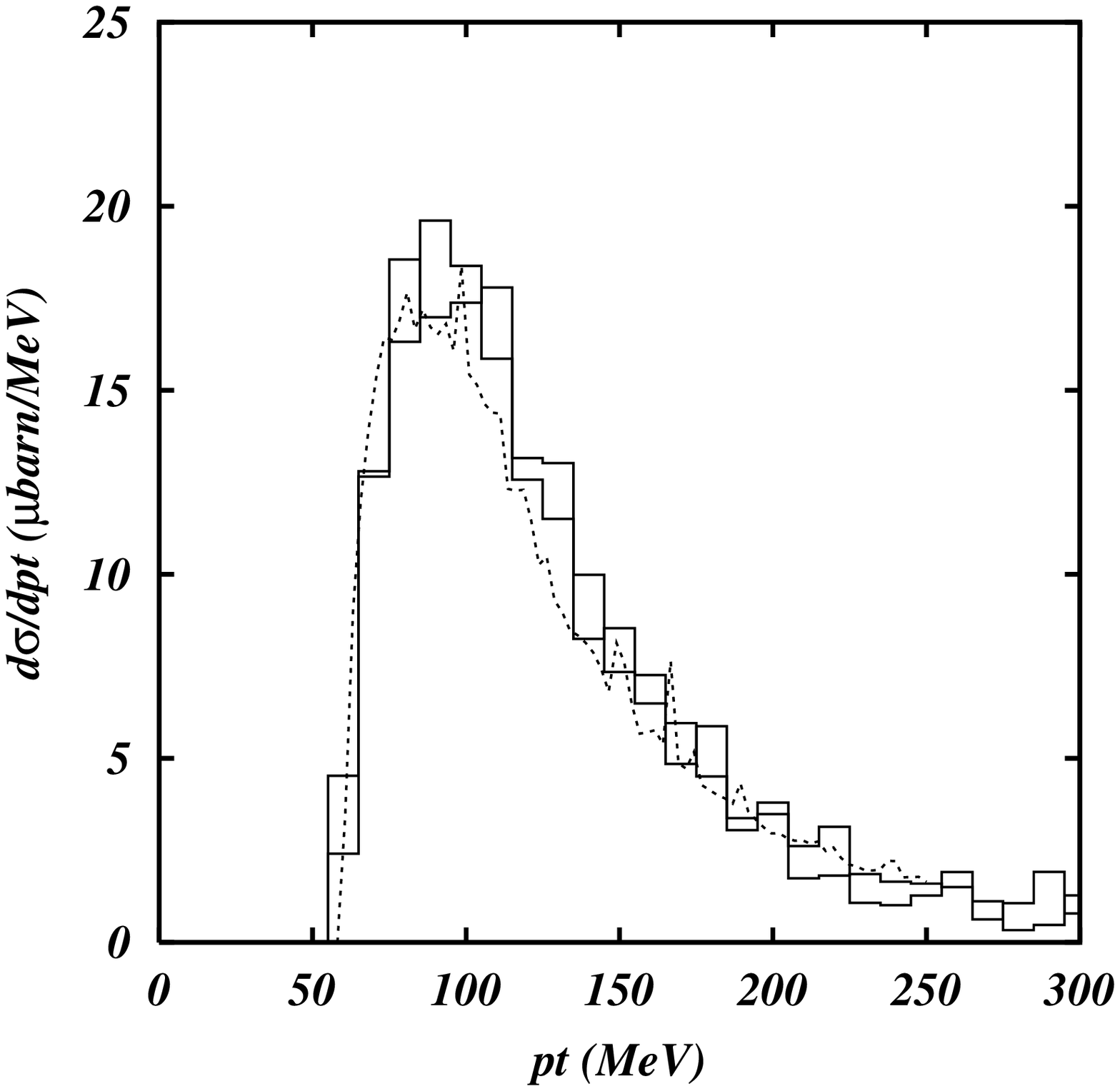}
\caption{Energy and transverse momentum distribution are shown for the electron
and the positron. In lowest order QED the two distributions are identical
and the spread between the two is a measure of the uncertainties coming from
the Monte Carlo integration.
This is compared with a unrestricted cross section calculation (dotted line),
normalized to give the same total cross section.}
\label{fig:rhic:ept}
\end{figure}

\begin{figure}[tb]
     \centering
     \includegraphics[width=7cm]{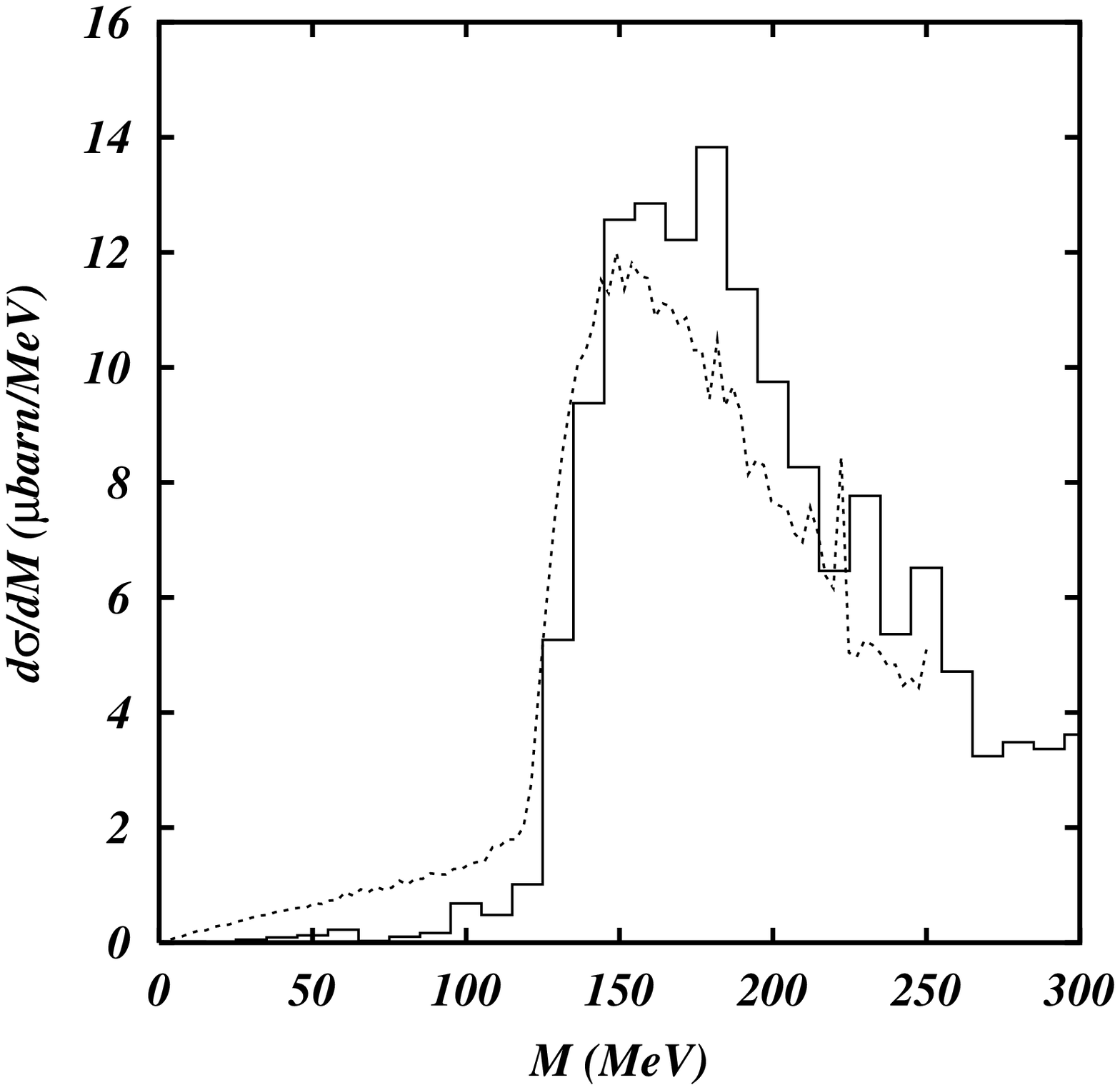}
     \includegraphics[width=7cm]{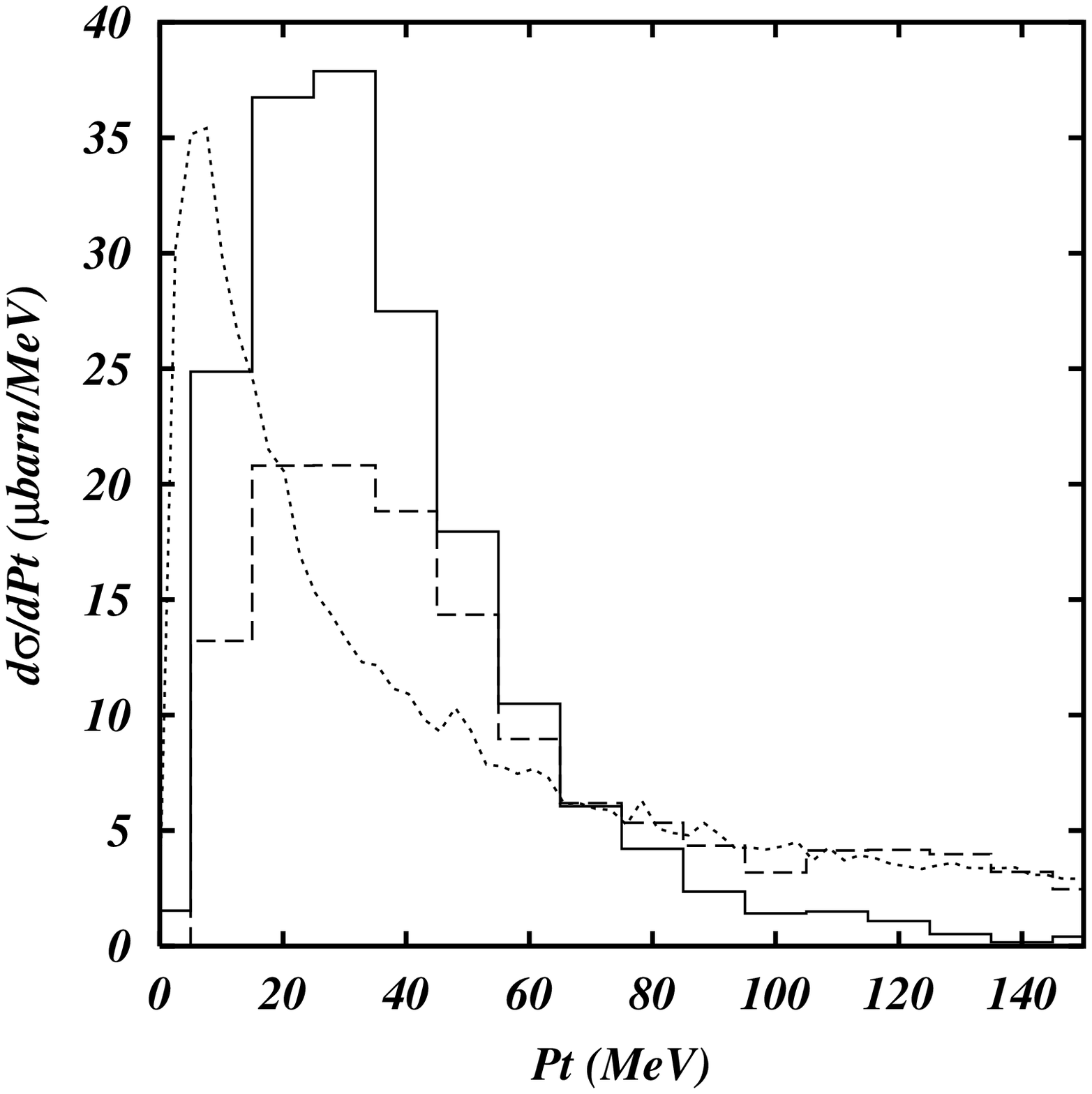}
\caption{The invariant mass and the transverse momentum of the $e^+e^-$ pair
are shown. Again the results are compared with the unrestricted cross section 
(dotted line). In addition
we have plotted the result of the restricted calculation without form factor
(dashed line). Both results have been normalized to give the same total cross 
section.}
\label{fig:rhic:pair}
\end{figure}

From this we get a total cross section for AuAu collisions at RHIC, including 
the restrictions $|p_t|>60$~MeV$/c$, $|y|<1.15$ of 2.30, 1.76, 1.43~mb, 
for $b_{min}=$ 13, 14 and 15fm, respectively.
In addition we have calculated a number of differential distributions, which
were also studied at STAR. The transverse momenta and energy distribution
of the electron and positron are shown in Fig.~\ref{fig:rhic:ept}. In lowest
order QED the distribution of electron and positron  are identical to each 
other. The difference between the two distributions can therefore be seen
as a measure of the accuracy the MC integration. 
One can look also at properties 
of the produced pair: The transverse momentum and the invariant mass of the 
pair are shown in 
Fig.~\ref{fig:rhic:pair}. We have
not shown the rapidity distribution, which we found to be more or less flat
over the allowed range. In all four diagrams we show also the
differential distributions of the unrestricted cross section. 
We have rescaled the data, so that the total cross section are the same in
both cases.
We see that the transverse momentum distribution and the energy 
distribution of the individual leptons are more or less identical in shape. The same
is also true for the invariant mass distribution, with the only exception that
the ``tail'' at low invariant masses is higher for the unrestricted 
distributions. The biggest effect is seen in the transverse momentum 
distribution of the pair. Here we also expect the effect of the small impact 
parameter (corresponding to larger transverse momenta of the photons) to be
largest. The slower fall-off at larger transverse momenta is most probably due
to the fact that no form factor was used in the calculation of this cross 
section, which should be visible at $P_t>80$~MeV$/c$. 
In order to investigate the effect of the nuclear form factor and in order
to understand the large difference between the cross section with and without
nuclear form factor, see Fig.~\ref{fig:total} above, we show in this plot
also the transverse momentum distribution of the pair for a calculation without
form factor. It can be seen that in this case the cross section gets 
sizeable contributions for $P_t>80$~MeV$/c$.


We have studied in addition the question, whether especially the distribution 
as a function of $P_t$ of the pair is sensitive to the form of $P_{GDR}(b)$
used in our approach. Using a more general approach, see appendix for details,
one expect that the next correction is of the form $P_{GDR}^3(b) \sim 1/b^6$
instead of $P_{GDR}^2(b) \sim 1/b^4$. In such a model the average impact 
parameter changes from about $\frac{8R_a}{3} \approx 19\mbox{~fm}$ to 
about $\frac{12R_a}{5} \approx 17\mbox{~fm}$, which is a small change compared
to the Compton wavelength of the electron (400~fm), but is still a 10\% 
reduction of this average impact parameter. We have therefore studied the 
distribution of electron and positron for this restriction. As expected the
differential distributions are found to be the same within the uncertainties
of the Monte Carlo approach used. 

\begin{figure}[tb]
     \centering
     \includegraphics[width=7cm]{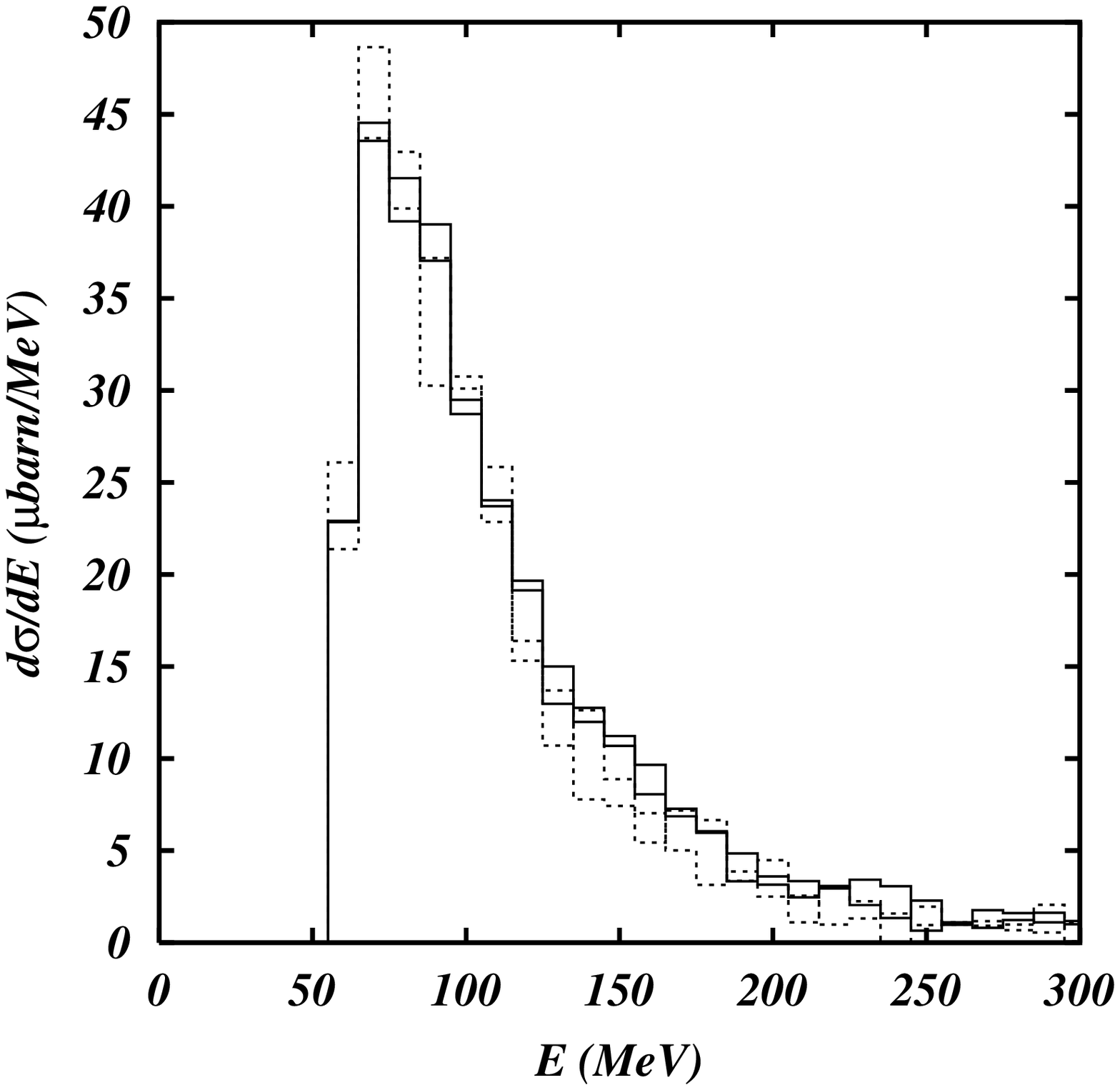}
     \includegraphics[width=7cm]{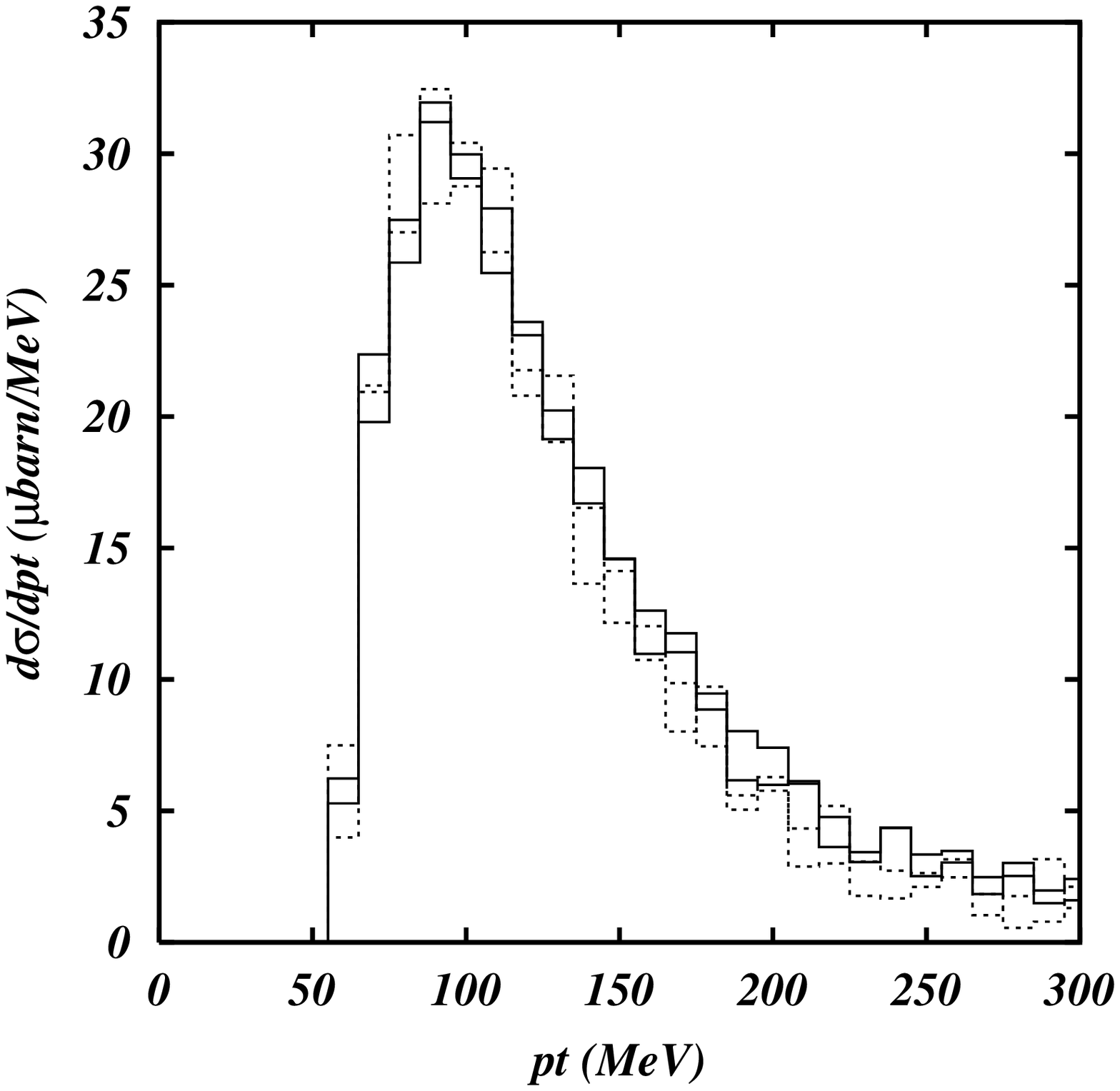}
\caption{Energy and transverse momentum distribution are shown for the two 
leptons for PbPb collisions at the LHC (solid line). This is compared with 
the rescaled spectrum of AuAu collisions at RHIC (dotted line).}
\label{fig:lhc:ept}
\end{figure}

\begin{figure}[tb]
     \centering
     \includegraphics[width=7cm]{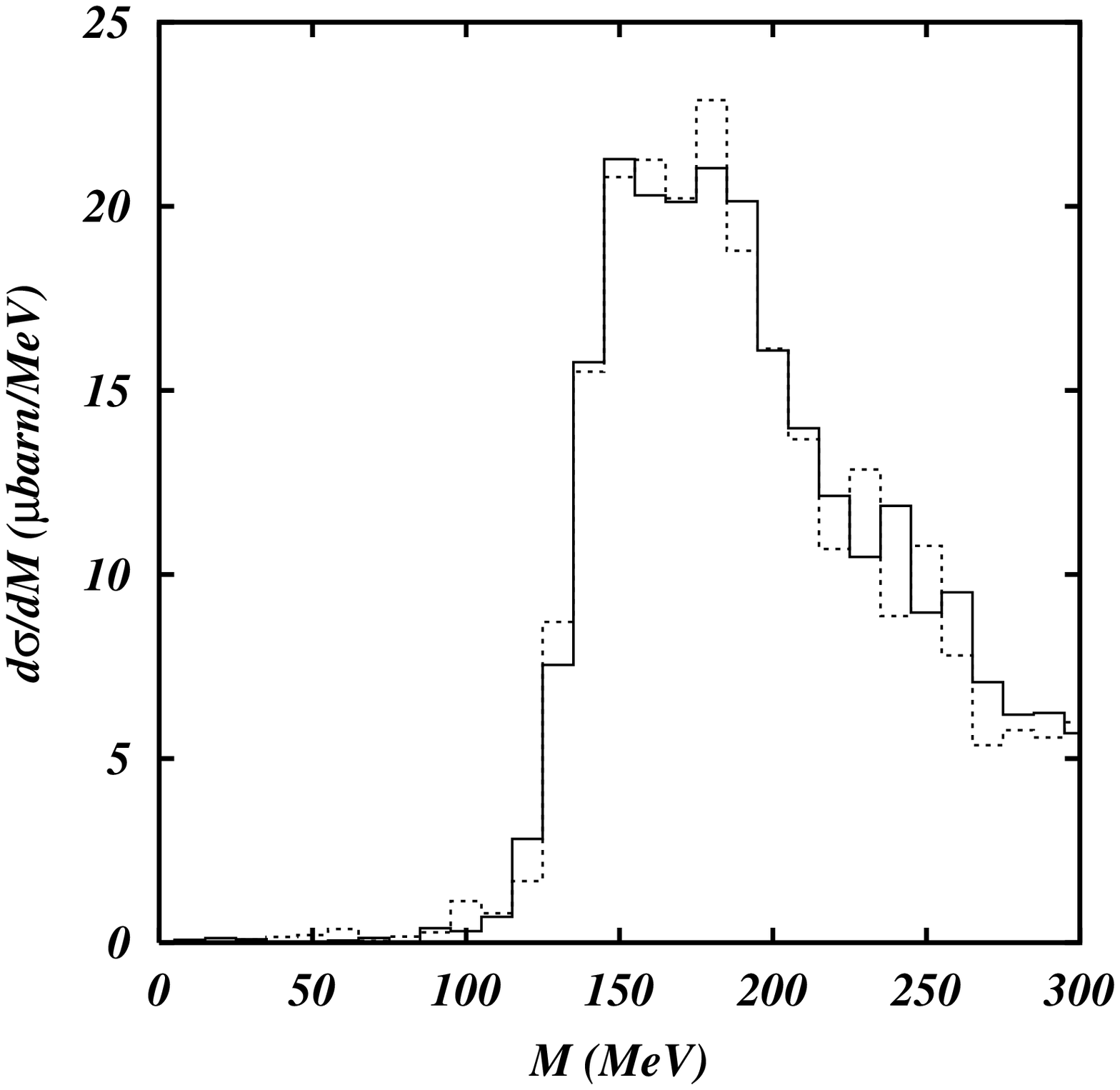}
     \includegraphics[width=7cm]{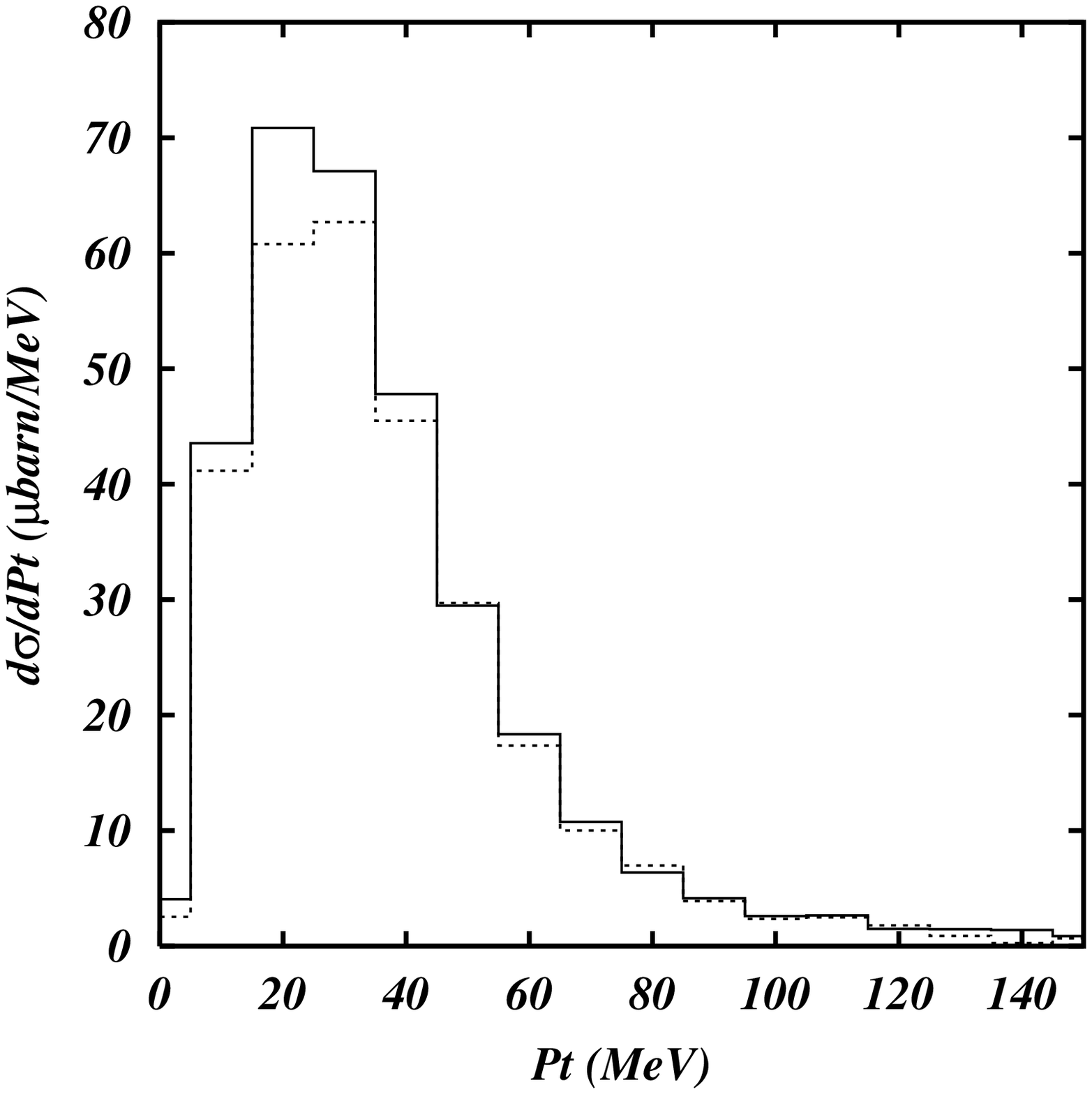}
\caption{The invariant mass and the transverse momentum of the $e^+e^-$ pair
is shown for PbPb collisions at the LHC (solid line).
The results are compared with the rescaled
spectrum of AuAu collisions at RHIC (dotted line).}
\label{fig:lhc:pair}
\end{figure}

As an outlook for future experiments we are showing results and 
distributions one might expect to see for PbPb collisions at the LHC.
Using the same kinematical restrictions as for the STAR experiment, the
results are shown in Fig.~\ref{fig:lhc:ept} and~\ref{fig:lhc:pair}. 

As a rather
optimistic estimate we have calculated also the differential cross 
section for a kinematical range of $|p_t|>2.6$~MeV$/c$ and $|y|<1.5$, where ALICE
will be able to detect the electrons with its Internal Trigger System (ITS), 
even though it will not be able to measure energies or momenta.


\begin{figure}[tb]
     \centering
     \includegraphics[width=7cm]{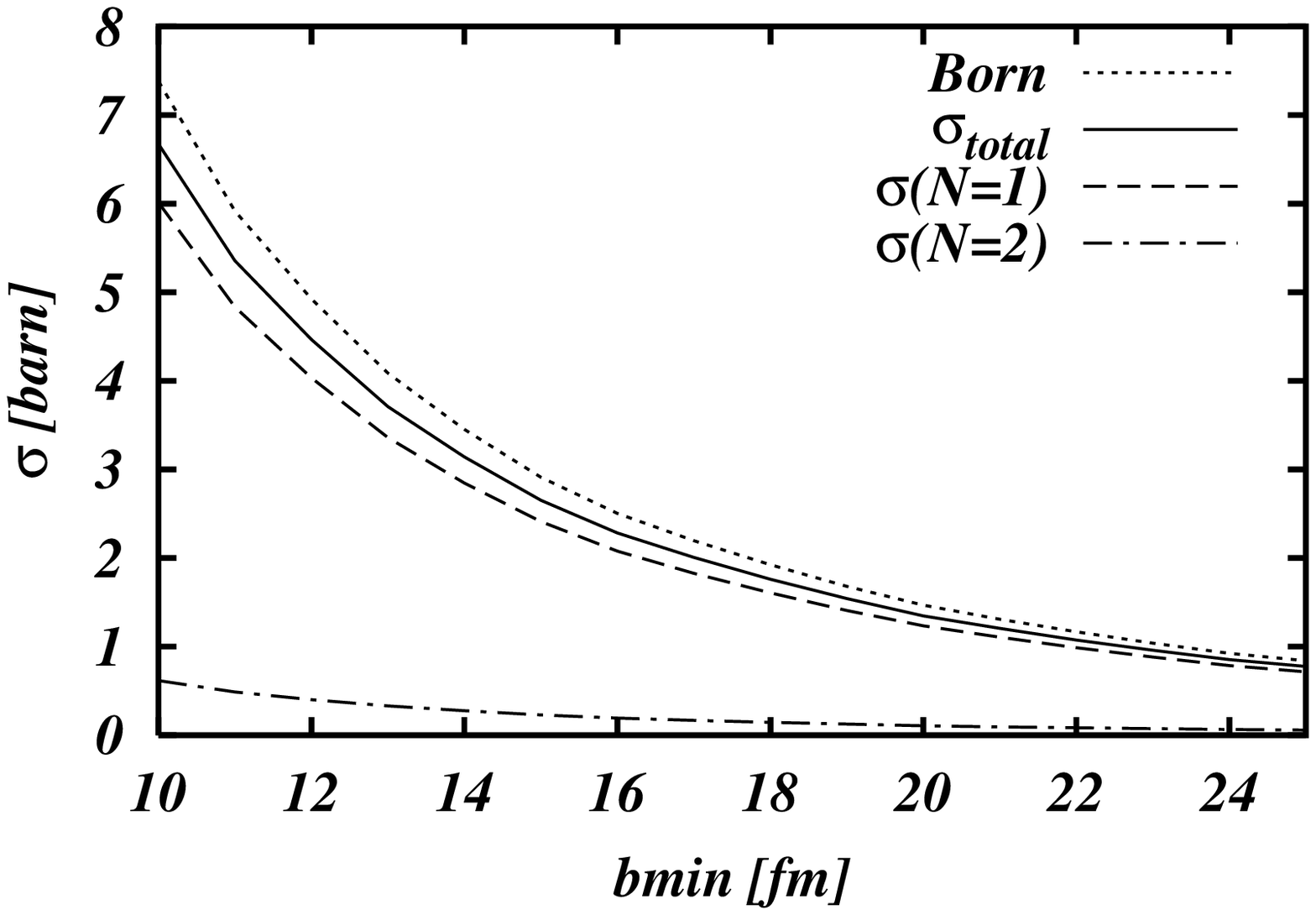}
\caption{The different cross sections for single and multiple pair production
are shown for the kinematical conditions at ALICE. The cross section is shown
as a function of $b_{min}$, the minimal impact parameter. See the text for
further details of the different cross sections.}
\label{fig:multi}
\end{figure}

With these kinematical conditions we can study the question, whether 
ALICE will be able to see multiple
pairs produced in a single collisions. For this we calculate the impact 
parameter dependent cross section under the kinematical conditions. One finds
that for impact parameters close to $b_{min}$ $P(b)\approx 20\%$. Following
\cite{HenckenTB95a,AlscherHT97,Guclu95,BaurHTS01} we use a Poisson distribution
as a good approximation for the $N$ pair production probability
\begin{equation}
  P(N,b) = \frac{P(b)^N}{N!} \exp(-P(b))
\end{equation}
Multiplying with $P^2_{GDR}(b)$ and integrating over $b$ we get the cross section
for one, two, \dots pair production. The result as a function of $b_{min}$ is
shown in Fig.~\ref{fig:multi} together with the total cross section 
\begin{equation}
  \sigma_{total} = \sum_{N=1}^{\infty} \sigma(N)
\end{equation}
The cross section one would get from the Born cross section can be interpreted
as a ``multiplicity'' cross section
\begin{equation}
  \sigma_{Born} = \int d^2b P^2(GDR,b) P(b) = \sum_{N=1}^{\infty} \sigma(N)
\end{equation}
and would be relevant in order to calculate the number of pairs produced
(in contrast
to the number of events). One can see that about 10\% of all events are
multiple pair production events and accordingly also about 10\% of all pairs
are produced in a multiple pair production process. This shows that at
ALICE one should be able to detect and study multiple pair production.

We have investigated a similar question also for RHIC using as an estimate for
a possible range $p_t>50$~MeV/$c$ and $2.5<y<4.0$. Unfortunately the 
probability for pair production under these conditions is only of the order
of a few permille and therefore the multiple
pair production cross section is less than one permille of the single pair 
production cross section, making such an investigation difficult.

\section{Discussion and Outlook}
We have calculated total and differential cross section of the pair production
process in ultraperipheral heavy ion collisions in lowest order QED for the
simultaneous electromagnetic excitation of both ions. We have seen that the
most sensitive quantity is the transverse momentum distribution, which differs
mostly from the distribution of the unrestricted cross section.
As the comparison will show \cite{Klein04,Meissner:2003me} our results were 
found to be in good agreement with the experimental results.
On the other hand only about 50 events were found at STAR, so the 
overall statistics is not very good. Additional runs might give better 
statistics. Still our analysis shows that the data at the moment give no
sign that higher order Coulomb effects are large for pairs produced with
these large transverse momenta \cite{Brodsky:1968rd}. Such higher order 
Coulomb effects would
lead most likely to an asymmetry of the electron and positron distribution
especially for the transverse momentum distribution. At the moment however
no calculation exists, which describes consistently the effects of the
strong Coulomb fields of both pairs on the pair production process at 
small impact parameter. The experimental condition of RHIC does not allow
to look for multiple pair production effects as the probability for pair
production with these conditions is well below one.

As already mentioned above the transverse momentum cut of $p_t>65$~MeV/$c$ at 
STAR was only possible due to a reduced magnetic field. There are currently
plans to use even lower magnetic fields and also making use of other detectors
within STAR \cite{SpencerPC} in order to extend the measurements both to
smaller transverse momenta and to larger rapidities. It remains to be
seen, whether the new phenomenon of  multiple pair production, will then
be detected. On the other hand with the low transverse momentum cutoff of
the ITS at ALICE/LHC, about 10\% of all pair production events are going to be
multiple pair production processes, therefore one should expect that this new
phenomena will be observed easily there. The fact that no kinematical 
information and also no particle identification is possible at ALICE, will make
such a measurement still a challenge.

\section{Acknowledgement}
The authors would like to thank Spencer R. Klein and Vladimir B. Morozov for
interesting discussions and collaborations on this subject and (S. K.) 
for critical reading of the manuscript. Discussions
with Serguei Sadovsky and Yuri Kharlov have been important in understanding
the possibilities of detecting the pair production at LHC.

\section{Appendix}
Throughout our calculations we have assumed that only the GDR excitation
is relevant for the triggering and therefore a simplified dependence on $b$,
see Eq.~(\ref{eq:pgdrb2}) has been used. In this appendix we want to show
that this is not a real limitation, but that other impact parameter 
dependencies can be treated as well. For example assuming that the higher
resonant states of the GDR are excited through a Poisson process, we would
need to replace $P_{GDR}(b)$ by $1-\exp(-P_{GDR}(b))$. 
In general we assume that
the relevant $P_{A\rightarrow A^* \rightarrow X + xn}(b)$ can be expressed as
a series of inverse powers of $b$
\begin{equation}
P_{A\rightarrow A^* \rightarrow X + xn}(b) = \sum_{n=0}^\infty 
\frac{S_n}{b^n}
\end{equation}
Using this in the expression for $d^6\sigma_{e^+e^-,2GDR}/(d^3p_+d^3p_-)$, see
Eq.~(\ref{eq:sigsb}), we need to calculate generalizations of $I(z)$, Eq.~(\ref{eq:iqb}) of the form
\begin{equation}
I_n(z) := \int_{z}^{\infty} \frac{dx}{x^n} J_0(x)
\end{equation}
with $I_3(z)$ corresponding to the one  used in our calculations.

For the calculation of these integrals, we first use the following recursion
relation 
\begin{equation}
I_n(z) = \frac{J_0(z)}{(n-1) z^{n-1}} - \frac{J_1(z)}{(n-1)^2 z^{n-2}}
- \frac{1}{(n-1)^2} I_{n-2}(z)
\label{eq:recursion}
\end{equation}
This relation can be easily derived by partial integration and by
using the well-known recursion relations between the Bessel functions
$J_n(z)$ \cite{AbramowitzS64}. By repeated application of
Eq.~(\ref{eq:recursion}) every $I_n$ with odd and even $n$ can be
reduced to the starting values $I_1(z)$ and $I_0(z)$, respectively,
which are given in the literature \cite{AbramowitzS64}:
\begin{equation}
  I_0(z) = 1 - z \ {}_1F_2\left(\frac{1}{2}; 1, \frac{3}{2}; -
  \frac{z^2}{4}\right) 
\end{equation}
and
\begin{equation}
  I_1(z) = \frac{z^2}{8} \ {}_2F_3\left(1,1;2,2,2; -\frac{z^2}{4}\right)
  - \ln \frac{z}{2} - \gamma 
\end{equation}
where $\gamma$ is the Euler constant.

These two expressions can be calculated easily by the rapidly
converging power series of the hypergeometric functions or by using
suited polynomial expressions given in the literature
\cite{AbramowitzS64}. We are here only interested in the case, where
$n=2m+1$ is an odd number, in which case the complete recursion
relation is given by
\begin{equation}
  I_{2m+1}(z) = \frac{(-1)^m}{m!^2 2^{2m+1}} \left\{
J_0(z) \sum_{s=1}^{m} s! (s-1)! \left(-\frac{4}{z^2} \right)^s -
\frac{z}{2} J_1(z) \sum_{s=1}^{m} (s-1)!^2 \left(-
  \frac{4}{z^2}\right)^s + I_1(z) \right\}.
\end{equation}
Moreover this equation can be further simplified, by splitting off the
terms singular at $z=0$
(the principal part of the Laurent expansion in $z$),
by using the power series for the Bessel functions:
\begin{equation}
  J_m(z) = \left(\frac{z}{2}\right)^m \sum_{k=0}^{\infty} \frac{(-z^2/4)^k}{k!(k+m)!}
\end{equation}
and by rearranging the resulting sum. After some straightforward
algebra we thus obtain the compact expression for $m\ge 0$
\begin{eqnarray}
I_{2m+1}(z) &=& \frac{(-1)^m}{2^{2m+3} (m+1)!^2} z^2 
\ {}_2F_3\left(1,1;2,2+m,2+m; -\frac{z^2}{4}\right) \\
& &- \frac{(-1)^m}{2^{2m} (m)!^2} \left( \ln \frac{z}{2} + \gamma \right)
+ \sum_{k=0}^{m} a_{m,k} z^{-2k}
\label{eq:I2mp1}
\end{eqnarray}
where
\begin{equation}
a_{m,0} = \frac{(-1)^{m}}{2^{2m} m!^2 } \sum_{s=1}^m \frac{1}{s},\quad
a_{m,k} = \frac{(-1)^{m-k}}{2^{2(m-k)+1} (m-k)!^2 k}\quad k\ge 1
\label{eq:amk}
\end{equation}
For $m=1$ we get the explicit expression
\begin{equation}
I_3(z) = \frac{1}{2 z^2} + \frac{1}{4} \left( \ln\frac{z}{2} + \gamma -1 
 \right) - \frac{1}{128} z^2
\ {}_2F_{3}(1,1;2,3,3;-z^2/4)
\end{equation}
and similarly for $m=2$. Again the hypergeometric
function in Eq.~(\ref{eq:I2mp1}) can well be calculated numerically by
its power series:
\begin{equation}
\ {}_2F_3\left(1,1;2,2+m,2+m; -\frac{z^2}{4}\right) = (m+1)! ^2
\left( -\frac{4}{z^2} \right) \sum_{k=1}^\infty
\frac{\left(-\frac{z^2}{4}\right)^k}{k (k+m)!^2}.
\end{equation}
We use this rapidly converging power series in our numerical calculations.
For the case where $n$ is an even number, the same approach can be
used. For completeness we only give here the final result:
\begin{equation}
  I_{2m}(z) = \frac{1}{2^{2m-1} (2m-1)}
  \ {}_1F_2\left(-m+\frac{1}{2};-m+\frac{3}{2},1;-\frac{z^2}{4}\right) +
  \frac{(-1)^m 2^{2m} m!^2}{(2m)!^2}.
\end{equation}

\end{document}